\documentclass[conf]{new-aiaa}
\usepackage[utf8]{inputenc}

\usepackage{graphicx}
\usepackage{subfig}
\graphicspath{ {./figures/} }

\usepackage{amsmath}
\usepackage{amssymb}
\usepackage{mathtools}

\usepackage{array}
\usepackage{multirow}
\usepackage{dcolumn}

\pdfoutput=1

\title{Convolutional Neural Network for Transition Modeling \\ Based on Linear Stability Theory}

\author{Muhammad I. Zafar}
\author{Heng Xiao\footnote{Corresponding author: hengxiao@vt.edu}} 
\affil{Kevin T. Crofton Department of Aerospace and Ocean Engineering, Virginia Tech, Blacksburg, VA 24061, USA}
\author{Meelan M. Choudhari} 
\author{Fei Li} 
\author{Chau-Lyan Chang} 
\affil{NASA Langley Research Center, Hampton, VA 23681, USA}
\author{Pedro Paredes} 
\author{Balaji Venkatachari} 
\affil{National Institute of Aerospace, Hampton, VA 23666, USA}

\begin{document}
\maketitle
\begin{abstract}
Transition prediction is an important aspect of aerodynamic design because of its impact on skin friction and potential coupling with flow separation characteristics. Traditionally, the modeling of transition has relied on correlation-based empirical formulas based on integral quantities such as the shape factor of the boundary layer. However, in many applications of computational fluid dynamics, the shape factor is not straightforwardly available or not well-defined. We propose using the complete velocity profile along with other quantities (e.g., frequency, Reynolds number) to predict the perturbation amplification factor. While this can be achieved with regression models based on a classical fully connected neural network, such a model can be computationally more demanding. We propose a novel convolutional neural network inspired by the underlying physics as described by the stability equations. Specifically, convolutional layers are first used to extract integral quantities from the velocity profiles, and then fully connected layers are used to map the extracted integral quantities, along with frequency and Reynolds number, to the output (amplification ratio). Numerical tests on classical boundary layers clearly demonstrate the merits of the proposed method. More importantly, we demonstrate that, for Tollmien-Schlichting instabilities in two-dimensional, low-speed boundary layers, the proposed network encodes information in the boundary layer profiles into an integral quantity that is strongly correlated to a well-known, physically defined parameter -- the shape factor.
\end{abstract}

\section{Introduction}
Laminar-turbulent transition of boundary-layer flows can have a strong impact on the performance of flight vehicles because of their influence on the surface skin friction and aerodynamic heating. Therefore, transition prediction is a key issue for the design of next-generation aerospace configurations. Indeed, according to the CFD Vision 2030 Study~\citep{green2005}, the most critical area in computational fluid dynamics (CFD) simulation capability that will remain a pacing item for the foreseeable future is the ability to adequately predict viscous flows involving transition-to-turbulence and/or flow separation.

Under the benign disturbance environment in flight, boundary-layer transition is often initiated by the amplification of linearly unstable eigenmodes of the laminar boundary layer. For 2D and weakly 3D boundary layers developing over a nominally smooth surface, the dominant instability mechanisms correspond to streamwise propagating Tollmien-Schlichting (TS) waves at subsonic speeds, oblique first-mode disturbances at supersonic edge Mach numbers, and again planar waves of the second mode, i.e., Mack mode type in hypersonic flows.  Additionally, centrifugal instabilities in the form of G\"{o}rtler vortices are known to influence the transition process over surfaces with significant regions of concave longitudinal curvature.  Finally, attachment line and crossflow instabilities come into play when the flow becomes three dimensional. 

The transition process begins with the generation of the instability waves via the interaction (i.e., receptivity) of the laminar boundary layer to its disturbance environment.  However, the onset of turbulence ensues only after the instability waves have gained sufficiently large amplitudes to undergo a sequence of nonlinear processes that culminates with the breakdown to turbulence.  Because the nonlinear breakdown tends to be relatively rapid, the slow amplification of the linear instability waves accounts for a majority of the laminar flow region preceding the onset of transition.  As a result, the linear amplification ratio, $e^N$, of the most amplified instability mode can often be used to predict the experimental trends in the transition location.  Prior work~\citep{ingen2008, bushnell1989} has shown that the N-factor values between 9 and 11 correlate with the transition locations measured in a broad class of boundary layer flows. 

Direct computations of boundary layer stability place rather stringent demands on the accuracy of mean flow calculations, much more so in comparison with that required for the prediction of aerodynamic metrics such as the skin friction drag.  In addition, the solution to the eigenvalue problem associated with the discretized version of the linear stability equations incurs a significant computational cost.  Furthermore, due to the complex nature of the eigenvalue spectra and their sensitivity to both input parameters and the numerical discretization, stability computations are difficult to automate, and also require significant user expertise into the details of the hydrodynamic stability theory.  Consequently, the task of transition prediction based on the N-factor methods has been a specialist’s domain, and is often performed as a post processing step that follows the computation of the laminar boundary layer over the flow configuration of interest. Implicit in this post processing approach is the assumption of a weak coupling between the transition location and the basic state computation.

The weak coupling assumption may be justified for fully attached boundary layer flows such as aircraft wings at the cruise condition.  However, a number of technological applications, such as high-lift systems~\citep{kusunose1991}, rotorcraft~\citep{sheng2018}, and other configurations involving flow separation, entail a strong viscous-inviscid interaction, requiring an iterative prediction approach that reflects the stronger coupling between transition and the overall flow field. To that end, the CFD Vision 2030 Study has called for transition prediction methods that can be fully integrated into the overall process of aerodynamic prediction.  In the past, several attempts have been made to simplify the application of the N-factor methods in the engineering environment, ranging from analytical but potentially complex data fits~\citep{ingen2008, drela2018, perraud2016} to numerical, table-lookup procedures based on a prior database of stability results~\citep{dagenhart1981, stock1989, gaster1995, langlois2012, krumbein2008, rajnarayan2013, begou2017, pinna2018}.  In recent work, such empirical fits have also been incorporated into  CFD integrated transition models based on auxiliary transport equations, such as the amplification factor transport model~\citep{coder2014}.

In the above-mentioned database-query techniques, a response surface model is developed in terms of a small number of scalar input parameters representing the combination of the global flow parameters, selected measures of boundary layer profiles, and the relevant disturbance characteristics such as frequency and wave-number parameters.  Almost universally, one or more shape factors of the boundary layer profiles have been used to encapsulate the complex dependence of  the disturbance amplification rates on the underlying mean flow.  This tends to limit the expressive power of the model as discussed by Crouch et al.~\citep{crouch2002}.  Secondarily, while such shape factors can be easily evaluated when the mean flow computation is based on the classical boundary layer theory, it is not easy to compute them in a consistent and accurate manner when Navier-Stokes codes are used for the basic state computation, especially for unstructured grid solutions.  Such situations arise rather commonly in high-speed applications, such as the flow past blunt nosed bodies, where the inviscid flow beyond the edge of the boundary layer includes nonzero vorticity as a result of vorticity generation at the curved shock.  The database methods can perform rather poorly in these cases as demonstrated by Paredes et al.~\citep{paredes2020}. 

Stability predictions based on artificial neural networks~\citep{crouch2002, fuller1997, danvin2018} allow additional features of the boundary layer profiles to be taken into account without sacrificing the computational efficiency and robustness of the conventional methods based on a previously computed database of amplification characteristics. The neural network methods can also be easily generalized to higher dimensional input features.  This allows the multiparameter dependence of the stability characteristics to be accounted for, whereas the conventional methods for database query do not scale very well as the number of independent parameters becomes significantly large.  Neural-network-based stability predictions for free-shear layer flows were first presented by Fuller et al.~\citep{fuller1997}.  However, a significant advance related to transition prediction was made by Crouch et al.~\citep{crouch2002}, who found that the expressivity of the model could be improved by augmenting the set of scalar variables used in conventional database methods via the slopes of the appropriately normalized velocity profiles at six equidistant points across the boundary layer.  The details of the neural network architecture used in that paper are somewhat limited; however, a feed-forward network based on fully connected hidden layers was used to approximate the maximum amplification rate among all unstable modes at any given station.  This maximum amplification rate was integrated along the airfoil surface to evaluate the N-factors. 

The selection of a smaller number of input features by Crouch et al.~\citep{crouch2002} was somewhat arbitrary and is unlikely to lend itself to other instability mechanisms without further modifications.  In the present work, we present an alternate approach based on  convolutional neural networks that can automatically learn a reduced-order representation of the boundary layer profiles in terms of a specified number of most significant features that can optimally predict the targeted linear stability characteristics across the training space.  As such, the proposed architecture can be easily adapted to predict the amplification characteristics of a broad range of very different instability mechanisms.  

The objective of this paper is to present a preliminary proof of concept to establish the potential of convolutional neural networks (CNN) to distill the latent features of the boundary layer profiles.  To that end, it is sufficient to consider the simplest case of Tollmien-Schlichting (TS) instability waves in two-dimensional, incompressible boundary layers.  Because the architecture of the CNN is not related to the specific physics of the TS instability mechanism or to the flow geometry, the proposed neural network can be easily generalized to other classes of instability waves.

The rest of the paper is organized as follows.
An overview of the methodology is presented in Section ~\ref{sec:methodology}, which includes a summary of modal stability analysis for two-dimensional, incompressible flows, followed by a description of the proposed architecture of a hybrid convolutional/fully-connected neural network.  Section~\ref{sec:results} presents the results based on the training and validation of the proposed network architecture, including an assessment of its generalization capability for real world applications.  In particular, we highlight the capability of the proposed network for encoding boundary layer profiles into integral parameters in an automatic, data-driven manner. Section~\ref{sec:conclude} concludes the paper.

\section{Methodology \label{sec:methodology}}
Using the $e^N$ method, the onset of laminar-turbulent transition is predicted to occur where the logarithmic amplification ratio~$N$ of the most amplified instability mode reaches an empirically defined critical value, denoted herein as $N=N_{tr}$. As described in section~II.A below, the logarithmic amplification ratio~$N$ can be computed by solving the governing equations based on linear stability theory. Thus, $N$ is a direct function of the laminar boundary layer profiles (e.g., $U$, $dU/dy$, and $d^2U/dy^2$), the flow Reynolds number $Re_{\theta}$ based on local momentum thickness of the boundary layer, and disturbance parameters such as the frequency of the instability wave $\omega$ and, for 3D disturbances, the spanwise wavenumber. The objective herein is to develop a surrogate transition model based on a neural network that would incorporate the physics of the transition phenomenon and predict the transition onset location without requiring the direct computations using linear stability theory. To that end, we propose a hybrid neural network (consisting of convolutional and fully-connected layers) that allows the relevant flow  information (boundary layer profiles and scalar disturbance characteristics) to be processed in a physically informed manner. The resulting regression models predict the local instability amplification rates, $\sigma$, for the relevant frequencies of the instability waves as they propagate through the laminar boundary layer. The $N$-factor curves can be computed by integrating the predicted $\sigma$ values, to allow the transition onset location to be predicted on the basis of the empirically defined critical value, $N_{tr}$. Moreover, as a baseline to compare the performance of the proposed neural network, we also consider the fully connected neural network architecture based on the idea proposed earlier by Crouch  et  al.~\citep{crouch2002}. In the first part of this section, we discuss the basis of the $e^N$ method and the linear stability theory. The transition models based on the proposed neural network and the fully connected neural network are outlined in the second part. The database used to train these neural-network-based transition models is discussed in the last part of this section.

\subsection{The $e^N$ Method \label{sec:eN}}
For simplicity, we outline the transition prediction procedure in the context of an incompressible, fully attached flow over a two-dimensional airfoil.
An orthogonal, body-fitted, curvilinear coordinate system ($s, n$) is introduced such that $s$ denotes the distance from the stagnation point, measured along the airfoil contour on either the suction or the pressure side of the airfoil, and $n$ represents the outward surface normal.  Consistent with the boundary-layer character of the basic state, both coordinates are normalized by a length scale comparable to the local thickness of the boundary layer, which is taken to be the local momentum thickness $\theta(s)$ without any loss of generality. The two-dimensional boundary layer flow over the airfoil is represented by the velocity field (U,V), where the velocities are normalized by a local velocity scale $U_e(s)$, taken to be the flow speed at the edge of the mean boundary layer.  The Reynolds number parameter based on the free-stream speed and the momentum thickness is denoted by $Re_{\theta}$.

We consider small-amplitude, time-harmonic, spatially evolving perturbations to the mean boundary layer flow of the form:
\begin{equation}
\begin{bmatrix}
u \\
v
\end{bmatrix}
= 
\begin{bmatrix}
\mathrm{U}(n) \\
\mathrm{V}(n)
\end{bmatrix}
\exp \left( i(\varphi(s)-\omega t) \right) \; ,
\end{equation}
where $i=\sqrt{-1}$ 
is the imaginary unit, $\omega \equiv 2 \pi f$ denotes the real valued disturbance frequency, $f$ is the frequency parameter in Hz, \textit{t} is the appropriately normalized time, and $d\varphi/ds = \alpha$ denotes the complex streamwise wavenumber.  Substituting the above normal mode ansatz into the linearized Navier-Stokes equations and neglecting the weak non-parallel effects associated with the O($Re^{-1}$) velocity component $\mathbf{V}$ and the slow streamwise evolution $\partial U/\partial s$ of the tangential velocity field, one obtains the quasiparallel form of the disturbance equations that must be solved with homogeneous boundary conditions for $\mathrm{U}$ and $\mathrm{V}$  at the surface ($n = 0$) as well as in the free stream ($n \rightarrow \infty$).  For an incompressible flow, the quasiparallel disturbance equations can be combined into a single equation for the wall-normal velocity perturbation, yielding an eigenvalue problem based on the well known Orr-Sommerfeld (OS) equation~\citep{drazin1981}:
\begin{equation}
    (\alpha U - \omega)(\mathrm{V}'' + \alpha^2 \mathrm{V})-\alpha U'' \mathrm{V} = (\mathrm{V}'''' - 2 \alpha^2 \mathrm{V}'' + \alpha^4)/(iRe_{\theta})
    \label{eq:OSeqn}
\end{equation}
along with homogeneous Dirichlet boundary conditions:
\begin{equation}
    \mathrm{V} (0) = \mathrm{V}'(0) = 0,
    \quad \text{and}  \quad
    \mathrm{V}(\infty) = \mathrm{V}'(\infty) = 0
    \label{eq:OSBC}
\end{equation}
where prime denotes the derivative with respect to the wall-normal coordinate \textit{n}. The solution to the eigenvalue problem (Eqs.~\ref{eq:OSeqn} and~\ref{eq:OSBC}) at a given station \textit{s} determines the local value of the complex streamwise wavenumber $\alpha$ as a function of the frequency parameter $\omega$. The local, streamwise amplification rate of a disturbance at frequency $\omega$ corresponds to $\sigma = - \text{Im}(\alpha(\omega, s ))$, where Im$(.)$ denotes the imaginary part of a complex quantity. Hence, the logarithmic amplification of the disturbance amplitude with respect to the neutral station, where the disturbance first begins to amplify, is given by:
\begin{equation}
N(\omega,s) = \int_{s_0}^{s} \sigma (\omega, \tilde{s}) d\tilde{s}
\end{equation}
where the subscript `0' denotes the neutral station. The $e^N$ method postulates that transition is likely to occur when 
the envelope N-factor, N$_e$(s)=sup(N($\omega$,s)), reaches the critical value of $N_e(s)=N_{tr}$. Here `sup' denotes the maximum over the frequency range of all unstable disturbances.  Values of $N_{tr}=9-11$ have been found to correlate with the onset of transition in a number of subsonic and supersonic flows~\citep{ingen2008, bushnell1989}.

\subsection{Convolutional Neural Network \label{sec:picnn}}
A neural network is a sequence of composite functions representing the mapping from an input vector $\mathbf{q}$ to output vector $\mathbf{y}$. Each member of the sequence is parameterized by the weight matrix $\mathbf{W}$ and bias vector $\mathbf{b}$, which can both be learned iteratively by using the available training data. This sequence of composite functions is arranged in the form of layers that consist of several neurons in general. For example, a neural network with one intermediate (hidden) layer $\mathbf{h}$ between the input layer ($\mathbf{q}$) and output layer $y$ may be represented by the following composite functional mapping:
\begin{equation}
    \mathbf{y} = \mathbf{W}^{(2)} \mathbf{h} + \mathbf{b}^{(2)} \quad \text{with} \quad \mathbf{h} = f [\mathbf{W}^{(1)} \mathbf{q} + \mathbf{b}^{(1)}], \; \text{or equivalently} \quad 
    \mathbf{y} = \mathbf{W}^{(2)} \left(f [\mathbf{W}^{(1)} \mathbf{q} + \mathbf{b}^{(1)}]\right) + \mathbf{b}^{(2)} ,
\end{equation}
where $f$ is an activation function, $\mathbf{W}^{(i)}$ and $\mathbf{b}^{(i)}$ represent the weight matrix and biases vector for the ${i}^\text{th}$ layer, respectively. Activation functions introduce the nonlinearity in the composite functions that enables them to represent arbitrarily complex functional relationships. Several different activation functions have been proposed for this purpose, such as the sigmoid function $ f(x) = 1/(1+e^{-x}) $ or the rectified linear unit (ReLU) $f(x) = \max(0,x)$. The training of the neural network consists of successive adjustments of the weights and the biases in order to minimize the squared error between the predicted and truth values of the output feature, i.e., the local amplification rate $\sigma$ in the present application. Neural networks with at least one hidden layer are \emph{universal approximators}, i.e., they can represent any continuous function on a compact domain to arbitrary accuracy, given a sufficient number of neurons in the hidden layer~\citep{hornik89multilayer}. 

\begin{figure}
    \centering
    \subfloat[]{\includegraphics[width=0.4\textwidth]{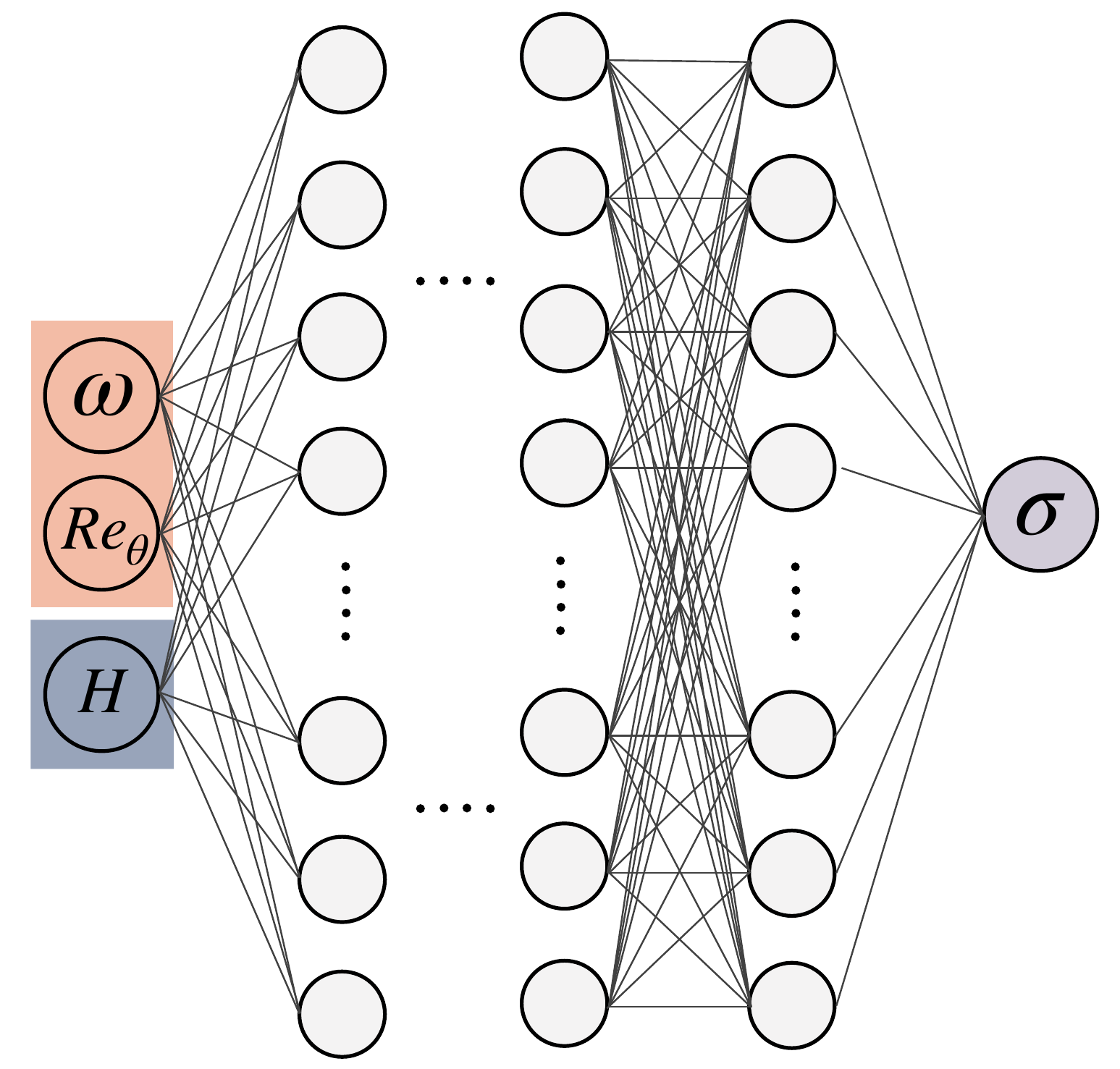}} \quad \quad \quad
    \subfloat[]{\includegraphics[width=0.4\textwidth]{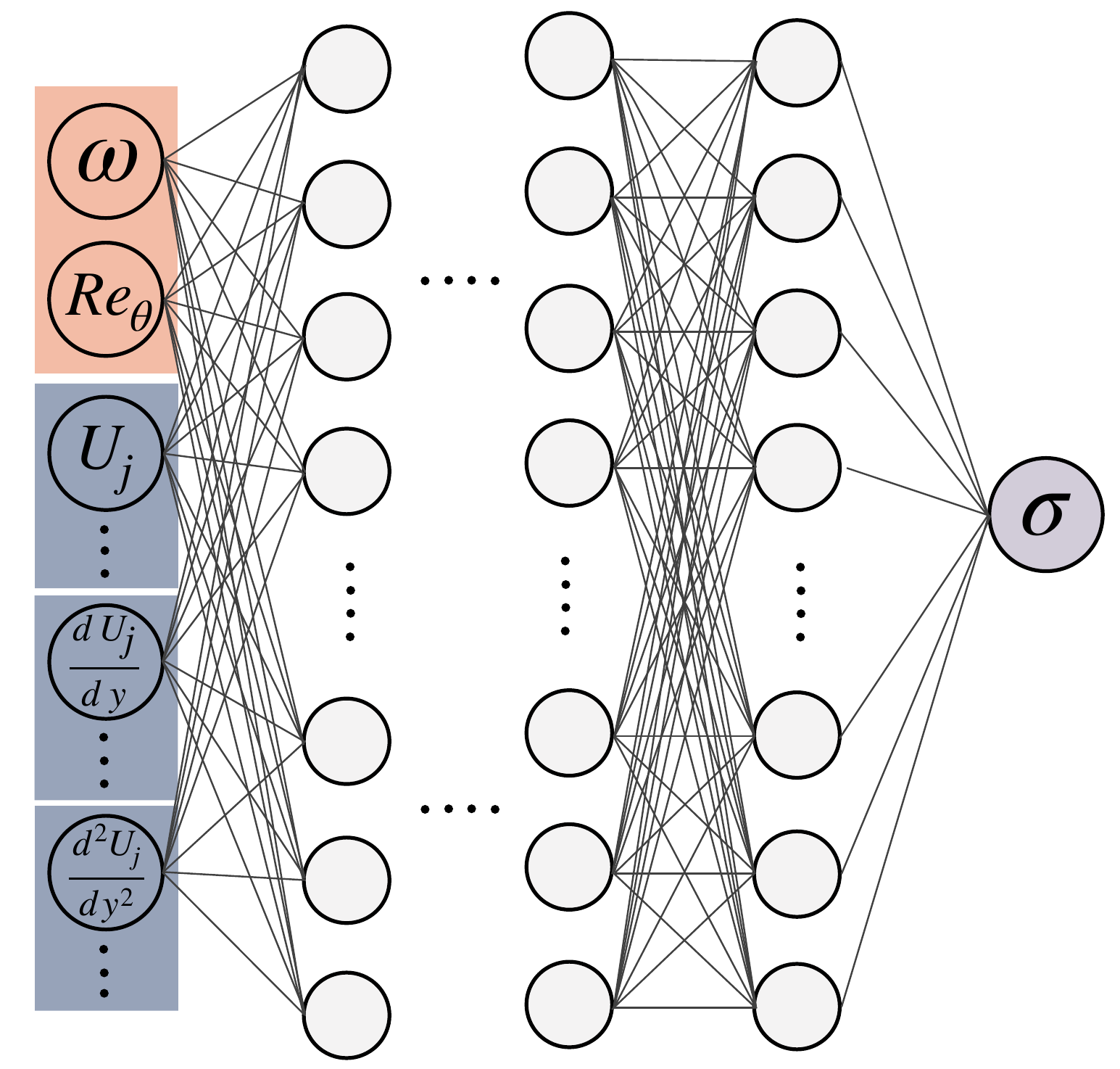}}
    \caption{Fully connected neural networks, \textbf{(a)} with input features including a scalar boundary layer parameter, shape factor $H$, \textbf{(b)} with boundary layer profiles as direct input features. Other scalar parameters in both architectures represents disturbance characteristics of instability waves, i.e., frequency ($\omega$) and Reynolds number ($Re_\theta$).}
     \label{fig:FCNN}
\end{figure}

In a fully connected neural network, each neuron in a given layer is connected to every neuron in the adjacent layer, yielding a generic connection pattern that makes no assumptions about the input features in the data. The schematic of a fully connected neural network with multiple input features is displayed in Fig.~\ref{fig:FCNN}. As mentioned in the Introduction, the existing models starting with Ref.~\citep{drela2018} have used analytical curve fits or other database query methods to predict the local amplification rate of the instability wave as a function of the disturbance frequency $\omega$, local Reynolds number $Re_{\theta}$, and a scalar parameter such as the shape factor $H$. 
A fully connected neural network such as that shown in Fig.~\ref{fig:FCNN}(a) provides a suitable architecture to achieve a similar functionality by using a neural network in place of the database interpolation or curve fitting. However, we note that the scalar boundary layer parameter $H$ does not appear directly in the governing equations for linear stability theory, and hence, is only indirectly related to the amplification characteristics of the instability modes. Moreover, the shape factor cannot be determined in a straightforward and/or accurate manner for several boundary layers, such as high speed flows over blunt nose configurations~\citep{paredes2020}. Therefore, it is desirable to introduce the boundary layer profiles directly into the predictive model, which is \emph{physically more consistent} to the underlying linear stability equation. Crouch et al.~\citep{crouch2002} presented a model of this type by including a coarse representation of the mean velocity profiles as part of the input for the fully connected neural network. Whereas they used the velocity and its first derivative at just six equidistant points, Fig.~\ref{fig:FCNN}(b) presents a model architecture based on a well resolved representation of the boundary layer profiles (i.e., velocity profile $U$ and its derivatives $dU/dy$ and $d^2U/dy^2$) as input features for the fully connected neural network. However, by using a large number of parameters to characterize the boundary layer profiles, a model architecture of this type risks a potential misrepresentation of the dependence of instability growth rates on the two remaining physical parameters, namely, $\omega$ and $Re_{\theta}$.  The balance involving the number of input parameters becomes more lop-sided for high speed flows, since the boundary layer profiles for temperature or density are also required for a reliable prediction of the amplification rate $\sigma$.

A convolutional neural network (CNN)~\citep{bishop2006} is composed of a number of convolutional and pooling layers, which enable it to automatically extract the latent features of the input data, which are considered to be an ordered data structures. Specifically, the boundary layer profiles can now be considered to be an ordered array, in contrast to the fully connected neural network that makes no assumption about the ordering of basic state quantities across the input profile.  Furthermore, the CNN exploits two special attributes to learn efficiently with a smaller number of model parameters. First, each neuron in the convolutional layers has only local connections to the neurons in the previous layer, allowing it to develop a correlation with the neighboring neurons. Second, the convolutional kernel has translational invariance in terms of model parameters, which leads to a drastic reduction in the number of network parameters. These attributes allow the CNN to achieve a comparable predictive accuracy much more efficiently in terms of the training cost and/or the amount of data required for training ~\citep{wu2018}. 

\begin{figure}
    \centering
    \includegraphics[width=0.99\textwidth]{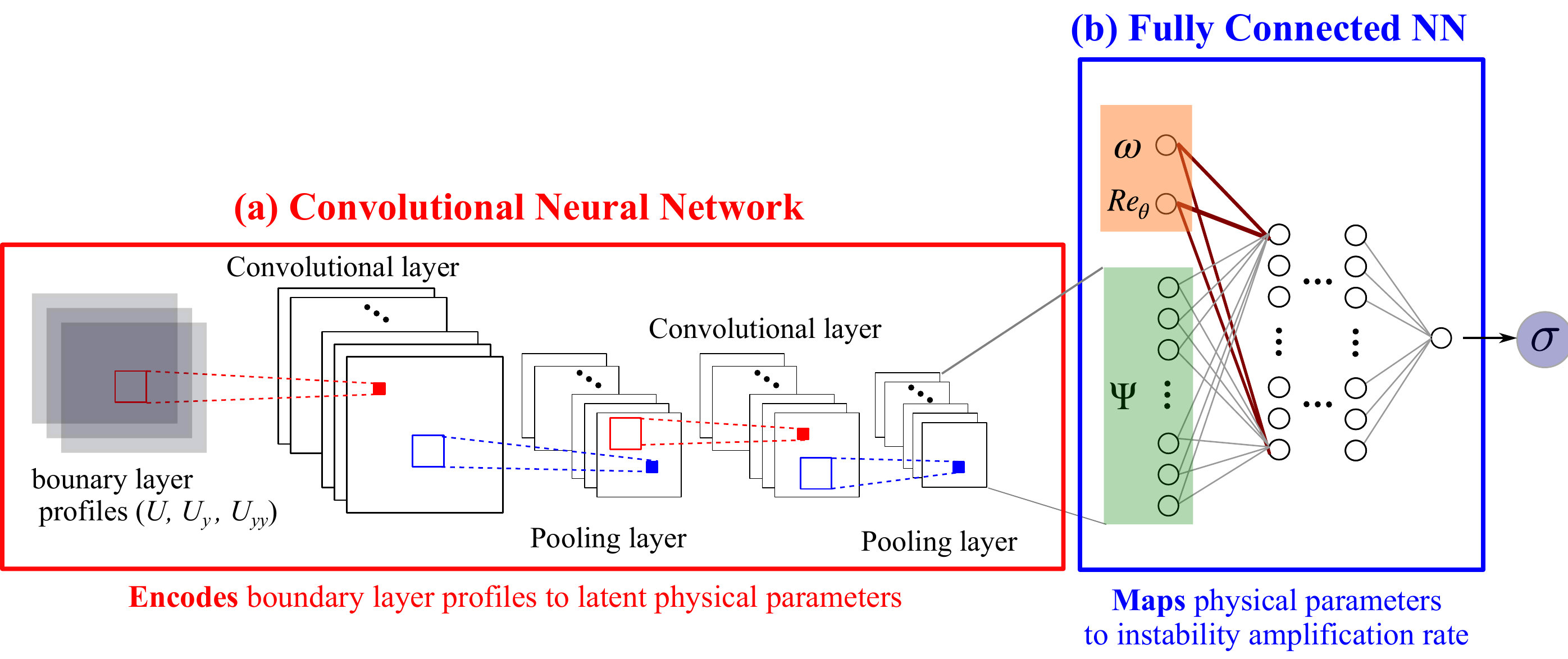}
    \caption{Proposed hybrid convolutional neural network architecture including (a) a regular CNN, which encodes the boundary layer profiles to a set of latent physical parameters $\Psi$ and (b) a fully connected neural network, which maps the CNN-extracted features $\Psi$ along with other physical parameters (frequency of the instability wave $\omega$ and Reynolds number $Re_\theta$) to the output (instability amplification rate $\sigma$).}
     \label{fig:PICNN}
\end{figure}
    
A schematic of the proposed hybrid neural network is presented in Fig.~\ref{fig:PICNN}. In this network architecture, the CNN first (Fig.~\ref{fig:PICNN}a) maps the boundary layer profiles to a specified number of latent features in a physically consistent manner while accounting for the spatial proximity of neighboring points across the boundary layer profiles. The encoding of the boundary layer profiles in the form of these latent features is denoted by the vector $\Psi$.  Following a preliminary assessment, the number of latent features in vector $\Psi$ was empirically chosen to be 8.  However, the results are not significantly sensitive to this parameter. Next, the revised set of input parameters comprising of the vector $\Psi$ and the remaining physical scalar parameters ($\omega$ and $Re_{\theta}$), is nonlinearly mapped through a fully connected neural network (Fig.~\ref{fig:PICNN}b) to yield the local instability amplification rate $\sigma$ as the final output. Observe that the dependence of $\sigma$  on the physical parameters $\omega$ and $Re_{\theta}$ is introduced into the network architecture in an explicit yet flexible manner. In particular, the relationship between the local instability amplification rate and the parameters $\omega$ and $Re_{\theta}$, along with the boundary layer profiles, is known from the linear stability theory (Eqs. 2,3), and that quantitative relation can be inferred via the training process.

The hyperparameters of the proposed neural network (Fig.~\ref{fig:PICNN}) and those of fully connected neural networks (Fig.~\ref{fig:FCNN}) have been empirically selected to yield an adequate complexity of the neural network model for learning all of the required information, without causing an overfitting of the training data. The list of primary hyperparameters includes the number of convolutional layers, the number of channels in each convolutional layer, the number of fully connected layers, the number of neurons in each fully connected layer, and the learning rate for the training of the model. A summary of the relevant model architectures is given in Table~\ref{tab:nn_details}, wherein each category of architecture has been labeled as A, B or $\text{C}_{1,2,3}$ for future reference. The number of neurons in each fully connected layer and the number of channels in each convolutional layer have also been listed for the respective architectures. 
The number of input channels for networks $C_1$, $C_2$ and $C_3$ can be varied to accommodate the desired number of boundary layer profiles that are to be used as input to the convolutional layers. Convolutional kernels of size $3 \times 1$ have been used to extract the latent features from the boundary layer profiles defined by 41 equidistant points along the wall-normal direction. In all networks, the Rectified Linear Unit (ReLU) is used as the activation function and the Adam optimization algorithm~\citep{adam2014} has been chosen to minimize the sum squared error during the training process. All of the neural network architectures considered herein have been implemented in the machine learning framework PyTorch.

\begin{table}[htbp]
    \centering
    \caption{Details of network architectures along with respective input features. Acronyms used: FC, fully connected; NN, neural network; NL, neurons in FC network layers; CH, number of channels in CNN layers. First fully connected layer of the $\text{C}_i$ ($i=1$, $2$, $3$) networks has (2+8) neurons, where the first 2 neurons correspond to the physical parameters ($\omega, Re_{\theta}$) and the 8 additional neurons correspond to a vector of parameters, $\Psi$, that encodes the information from boundary layer profiles. The $\text{C}_i^*$ ($i=1$, $2$, $3$) networks use only one scalar value for representing parameter $\Psi$, hence the first fully connected layer has (2+1) neurons.}
    \label{tab:nn_details}
    \begin{tabular}{m{0.06\textwidth} m{0.2\textwidth} m{0.24\textwidth} m{0.001\textwidth} m{0.26\textwidth} m{0.001\textwidth} m{0.08\textwidth}}
     \hline
     Network & Input features & Architecture Type & & Architecture & & Number of Parameters \\
     \hline
     $\text{A}$ & $ \omega, Re_{\theta}, H $ & Fig.~\ref{fig:FCNN}(a):~Fully Connected NN & & NL:~[3,96,96,96,96,96,96,1] & & 56,353 \\
     \hline
     $\text{B}$ & $ \omega, Re_{\theta}, U_j, \left. \frac{dU}{dy}\right\vert_j, \left. \frac{d^2U}{dy^2}\right\vert_j$ & Fig.~\ref{fig:FCNN}(b):~Fully Connected NN & & NL:~[125,96,96,96,96,96,96,1]  & & 57,261 \\
     \hline
    $\text{C}_1$ & $ \omega, Re_{\theta}, U_j, \left. \frac{dU}{dy}\right\vert_j, \left. \frac{d^2U}{dy^2}\right\vert_j$ & Fig.~\ref{fig:PICNN}:~Convolutional NN + Fully Connected NN & & CH:\enskip[3,6,8,4]\quad+ $\quad \quad \quad \quad \quad \quad$ NL:\enskip[2+8,96,96,96,96,96,96,1]  & & 57,337 \\
     \hline
     
     $\text{C}_2$ & $ \omega, Re_{\theta}, U_j, \left. \frac{dU}{dy}\right\vert_j $ & Fig.~\ref{fig:PICNN}: Convolutional NN + Fully Connected NN & & CH:\enskip[2,4,8,4]\quad+ $\quad \quad \quad \quad \quad \quad$ NL:\enskip[2+8,96,96,96,96,96,96,1]  & & 57,257 \\ \hline
     
     $\text{C}_3$ & $ \omega, Re_{\theta}, U_j$ & Fig.~\ref{fig:PICNN}: Convolutional NN + Fully Connected NN & & CH:\enskip[1,4,8,4]\quad+ $\quad \quad \quad \quad \quad \quad$ NL:\enskip[2+8,96,96,96,96,96,96,1] & & 57,245 \\ \hline
     
     $\text{C}_i^*$ i=1,2,3 & Same as $\text{C}_i$ above & Same as $\text{C}_i$ above & & Same as $\text{C}_i$ above, but with only 2+1 input features for the fully connected network & & 57,245 \\ \hline
    \end{tabular}
\end{table}

\subsection{Generation of training data \label{sec:gen_data}}
The training database is obtained by solving the Orr-Sommerfeld (OS) eigenvalue problem (Eq.~\ref{eq:OSBC}) for the Falkner-Skan family of self-similar boundary layer profiles over a wide range of pressure gradient parameter $\beta_H$ and local Reynolds number $Re_\theta(s)$.  It was generated by using the stability analysis software, LASTRAC, developed at the NASA Langley Research Center~\citep{chang2004}.  LASTRAC is a well-known software suite that has been extensively validated against existing benchmark data including direct numerical simulations.

The convolutional neural network model maps the complex dependence of the local instability amplification rate on the relevant disturbance characteristics and the mean flow parameters. The training database consists of a tabular listing of stability characteristics for the Falkner-Skan boundary layers, which includes a comprehensive, discrete sampling of the complex-valued local wavenumber $\alpha$ of the Tollmien-Schlichting (TS) instability wave as a function of the real-valued frequency of the wave $\omega$ and the mean flow parameters, which include the Hartree pressure gradient parameter $\beta_H$, or equivalently, the shape factor $H$ of the Falkner-Skan velocity profile, the local Reynolds number $Re_\theta$ based on the momentum thickness, and the velocity profile and its first- and second-order derivatives.

The Falkner-Skan group of boundary layers supports a single instability mode that corresponds to the viscous-inviscid interactive Tollmien-Schlichting (TS) waves by themselves ($\beta_H \ge 0$) or a combination of instability mechanisms involving the TS waves and the predominantly inviscid Rayleigh instabilities ($\beta_H < 0$).  All stability calculations were carried out for a compressible boundary layer flow with a vanishingly small Mach number of 0.00001, and a stagnation temperature of 311.11~K, along with an adiabatic thermal wall boundary condition and zero transpiration velocity at the surface.
All parameters included in the database are nondimensional. Lengths are scaled with respect to local momentum thickness, velocities with respect to the flow speed at the edge of the boundary layer, and temperature with respect to the local edge temperature. The database includes Hartree pressure gradient parameters in the range of $\beta_H \in [-0.1988,1]$, corresponding to the discrete values given by:
\begin{equation*}
\begin{split}
\beta_H = &[-0.1988, -0.19, -0.18, -0.16, -0.14, -0.12, -0.10, -0.075, -0.05, -0.025, \\
& \quad 0, 0.025, 0.05, 0.075, 0.1, 0.15, 0.2, 0.3, 0.4, 0.5, 0.6, 0.8, 1.0 ].
\end{split}
\end{equation*}
Because of a rapid change in the instability characteristics for $\beta_H<0$, especially as $\beta_H \to -0.1988$ (i.e., when the boundary layer is on the verge of separation), the sampling in $\beta_H$ is chosen to be denser at the negative values of $\beta_H$.  This bias in sampling may result in a bias in the learning of the neural network models toward the lower limit of the $\beta_H$ values. 
Stability computations were carried out for Reynolds numbers extending from just below the minimum critical Reynolds number (below which all disturbances are predicted to decay) up to $Re_s \equiv U_e s/\nu = 10^{10}$, where $\nu$ denotes the kinematic viscosity of the fluid.  The frequency range at each Reynolds number included the entire range of unstable disturbances as well as a modest range of stable disturbances in the vicinity of the neutral stability curve.  Because the parameter range covered multiple orders of magnitudes, a logarithmic increment was used along both axes.
Due to computational considerations, a quarter of the data points in the database were used, which amounts to approximately 400,000. The results are not influenced by the down-sampling of training data. The input parameters used to train and predict the local instability amplification rate correspond to a suitable subset of the various features listed in Table~\ref{tab:inputfeatures}. 
Boundary layer profiles include the velocity profile along with the first- and second-order derivatives, sampled at 41 equidistant points in order to resolve each profile. The scalar input features ($\text{q}_1: \omega$, $\text{q}_2: Re_{\theta}$ and $\text{q}_3: H$) have been scaled and shifted to the range of $[0, 1]$.  We note in passing that modified input features based on a logarithmic scale along the $Re_\theta$ and $\omega$ axes were also considered on the basis of the high Reynolds number asymptotic theory of Tollmien-Schlichting waves~\citep{Smith1979}, but no significant improvement in the testing error was noted. 

\begin{table}[htbp]
\caption{Input features for neural network models.}
 \label{tab:inputfeatures}
\centering
\begin{tabular}{ c l c }
\hline
Feature &	Definition & Expression \\
\hline
$q_1$ & Nondimensional frequency of the instability wave & $\omega$ \\
$q_2$ & Reynolds number based on edge velocity and momentum thickness & $Re_{\theta}$ \\
$q_3$ & Local value of velocity profile shape factor (derived parameter) & $H$ \\
$q_4$ & Velocity profile as a function of wall normal coordinate y  & $ U_j$, $j=1, 2, \dots, 41$ \\
$q_5$ & First order derivative of velocity profile & $\left. \frac{dU}{dy}\right\vert_j$, $j=1, 2, \dots, 41$ \\
$q_6$ & Second order derivative of velocity profile & $\left. \frac{d^2U}{dy^2}\right\vert_j$, $j=1, 2, \dots, 41$ \\
\hline
\end{tabular}
\end{table}

The present database has been generated as part of the NASA Langley Research Center's effort to use machine learning methods to enable robust, CFD-solver friendly models for boundary layer transition. This database will be made available in an electronic form to encourage the development of physics-based transition models that can be integrated with CFD flow solvers.

\section{Results \label{sec:results}}
In this section, we demonstrate the predictive performance of the proposed convolutional neural network (Fig.~\ref{fig:PICNN}) and compare it with the performance of the fully connected neural networks (Figs.~\ref{fig:FCNN}a and~\ref{fig:FCNN}b). In the first part of this section, the proposed neural network model is validated using the Falkner--Skan database (presented in Section~\ref{sec:gen_data}) by randomly splitting the data, with 90\% of the data points used for training and the remaining 10\% for validation. Both the training and validation datasets contain data from the entire range of the pressure gradient parameter $\beta_H$. We then assess the proposed model for interpolation and extrapolation cases by splitting the Falkner--Skan database into training and testing datasets based on the data corresponding to each pressure gradient parameter $\beta_H$. 

Finally, the capability of the network trained on the Falkner-Skan database (with self-similar boundary layers) to generalize its predictions to realistic flow configurations is evaluated by comparing the predictions of the neural network model with actual stability computations for those configurations.  Specifically, we consider two different airfoils with non-self-similar boundary layers for this purpose, namely, a symmetric 2D HSK airfoil~\citep{schetz1999} and an asymmetric NLF-0416 airfoil~\citep{somers1981, gopalarathnam2001}. The Reynolds number parameter based on the free-stream speed and the chord length of the airfoil is chosen to be $Re_c = 1.23\times10^6$ for the HSK airfoil and $Re_c=9\times10^6$ for the NLF-0416 airfoil. At the selected flow conditions, the boundary layer instability is dominated by the TS instabilities of interest. In all cases, the following metric corresponding to the percent error based on the Frobenius norm is used for the evaluation of the model throughout this paper:

\begin{equation}
    \epsilon_\sigma = 100 \times \frac{\|\mathbf{\sigma}_\text{truth} - \mathbf{\sigma}_\text{predicted}\|_F}{\|\mathbf{\sigma}_\text{truth} \|_F} 
    \label{error}
\end{equation}
where the Frobenius norm is defined as $\|X\|_F=\sqrt{\sum_{i}\left|X_i\right|^2}$. In the second part of this section, we analyse the feature learning and encoding capability of the convolutional neural network and how it makes the proposed model more robust and generalizable to other flow regimes. In the third part of this section, the potential advantages of the proposed model over the previously proposed model architecture~\citep{crouch2002} are analysed.

\subsection{Demonstration of Predictive Performance}
The predictive performance of the convolutional neural network (Fig.~\ref{fig:PICNN}) is first validated by using the Falkner--Skan database and is also compared with the performance of the fully connected network (Fig.~\ref{fig:FCNN}a) with scalar input features. In these cases, the training process utilized a randomly sampled subset of the available dataset, amounting to 90\% of the total data points. The validation is conducted by using the remaining 10\% of the data that were never seen by the neural network model during the training process.  The results presented in Table~\ref{tab:results} show that the proposed model demonstrates slightly improved predictive performance (Network C1, 0.41\%) as compared to that of the fully connected neural network with scalar input features (Network A, 0.58\%). Furthermore, the proposed neural network provides qualitatively similar results when the number of velocity profiles used as input features is varied from the velocity and its two derivatives (network $C_1$) to the velocity profile alone (network $C_3$). The data presented in Table~\ref{tab:results} indicate that the inclusion of the velocity derivatives decreases the validation error, but the improvement is rather small.  The small reduction in validation error is consistent with the fact that the derivative information is contained within the velocity profile itself, and therefore, we believe that the observed improvement is attributed to the well resolved yet finite sampling of the velocity profile.
Since the validation error for the neural network based on the velocity profiles alone (i.e., network $C_3$) is already small, the margin for improvement is rather limited, and the findings in Table~\ref{tab:results} confirm this expected behavior.  Given the small differences in validation error corresponding to the networks $C_1$ through $C_3$, all of the remaining assessments reported in this paper are based on a single set of input features, which is chosen to include all three boundary layer profiles.  Thus, the mapping sought by the neural network may be represented as:     

\[ 
\left(\omega, Re_{\theta}, U_j, \left. \frac{dU}{dy}\right\vert_j, \left. \frac{d^2U}{dy^2}\right\vert_j\right) \quad \longmapsto \quad \sigma ,
\quad \text{where} \quad j=1, 2, \dots, 41.
\]

\begin{table}[htbp]
    \centering
     \caption{Comparison of the validation error corresponding to neural network models from Table~\ref{tab:nn_details}.  Training and validation datasets correspond to a random 90\%-10\% split of the available database over the entire range of the pressure gradient parameter $\beta_H$.}
    \label{tab:results}
    \begin{tabular}{l l c}
     \hline
    Network & Input features & Validation error \\
     \hline
    A & $ \omega, Re_{\theta}, H $ & 0.58\% \\
    \hline
    $\text{C}_1$ & $ \omega, Re_{\theta}, U_j, \left. \frac{dU}{dy}\right\vert_j, \left. \frac{d^2U}{dy^2}\right\vert_j$ & 0.41\% \\
    $\text{C}_2$ & $ \omega, Re_{\theta}, U_j, \left. \frac{dU}{dy}\right\vert_j $  & 0.44\% \\ 
    $\text{C}_3$ & $ \omega, Re_{\theta}, U_j$ & 0.46\% \\ 
    \hline
    \end{tabular}
\end{table}

We now evaluate the performance of the proposed neural network for more challenging interpolation and extrapolation cases, where the Falkner--Skan database has been split for testing and training based on the Hartree pressure gradient parameter $\beta_H$. In these cases, data corresponding to a selected value of $\beta_H$ are reserved for testing while the data corresponding to the remaining values of $\beta_H$ are used for the training process. 
Interpolation cases were considered by isolating a single value of $\beta_H$ for evaluating the testing error and the corresponding results for $\beta_H = -0.1, 0.1$, and $0.5$ are given in Table IV.   The proposed neural network model is able to interpolate the remaining database at each of these three values of $\beta_H$ with a testing error of between 3\% and 5\%.  We also considered two additional cases based on testing corresponding to the two extremes of the $\beta_H$ range, namely, $\beta_H=-0.1988$ and $\beta_H=1.0$, which amount to an extrapolation from the training database.  As expected, the corresponding results in Table IV indicate significantly higher testing errors with the extrapolation in comparison with the testing errors for the three interpolation cases discussed above.  The testing error percentage toward the lower limit ($\beta_H=-0.1988$) is significantly lower (15.3\%) than that toward the higher limit of the range of pressure gradient parameters  (namely, 28.5\% error for $\beta_H = 1.0$). This disparity in testing errors may be due to the significantly denser sampling in $\beta_H$ toward the lower end of the range, indicating that the overall accuracy of the neural network could perhaps be improved by increasing the database size by including additional data at the higher values of $\beta_H$.  

\begin{table}[htbp]
\caption{Results for interpolation and extrapolation in the Hartree pressure gradient parameter. The testing dataset is comprised of data at a specified value of $\beta_H$, while the remaining data are used as the training dataset.}
\label{tab:cases}
\centering
\begin{tabular}{c c c}
\hline
Cases & Testing dataset & Testing error \\
\hline \hline
\multirow{3}{*}{Interpolation}   &   $\beta_H = -0.1$  &    \quad 3.97 \% \\ \cline{2-3}
                                                &   $\beta_H = 0.1$   &    \quad 3.02 \% \\ \cline{2-3}
                                                &   $\beta_H = 0.5$   &  \quad 4.68 \% \\ \hline \hline
\multirow{2}{*}{Extrapolation}  &   $\beta_H = -0.1988$  &    \quad 15.34 \% \\ \cline{2-3}
                                                &   $\beta_H = 1.0$    &    \quad 28.52 \% \\ \hline
\hline
\end{tabular}
\end{table}

\begin{figure}
    \centering
      \subfloat[HSK airfoil, $\alpha = 0^\circ$]{ \includegraphics[width=0.4\textwidth]{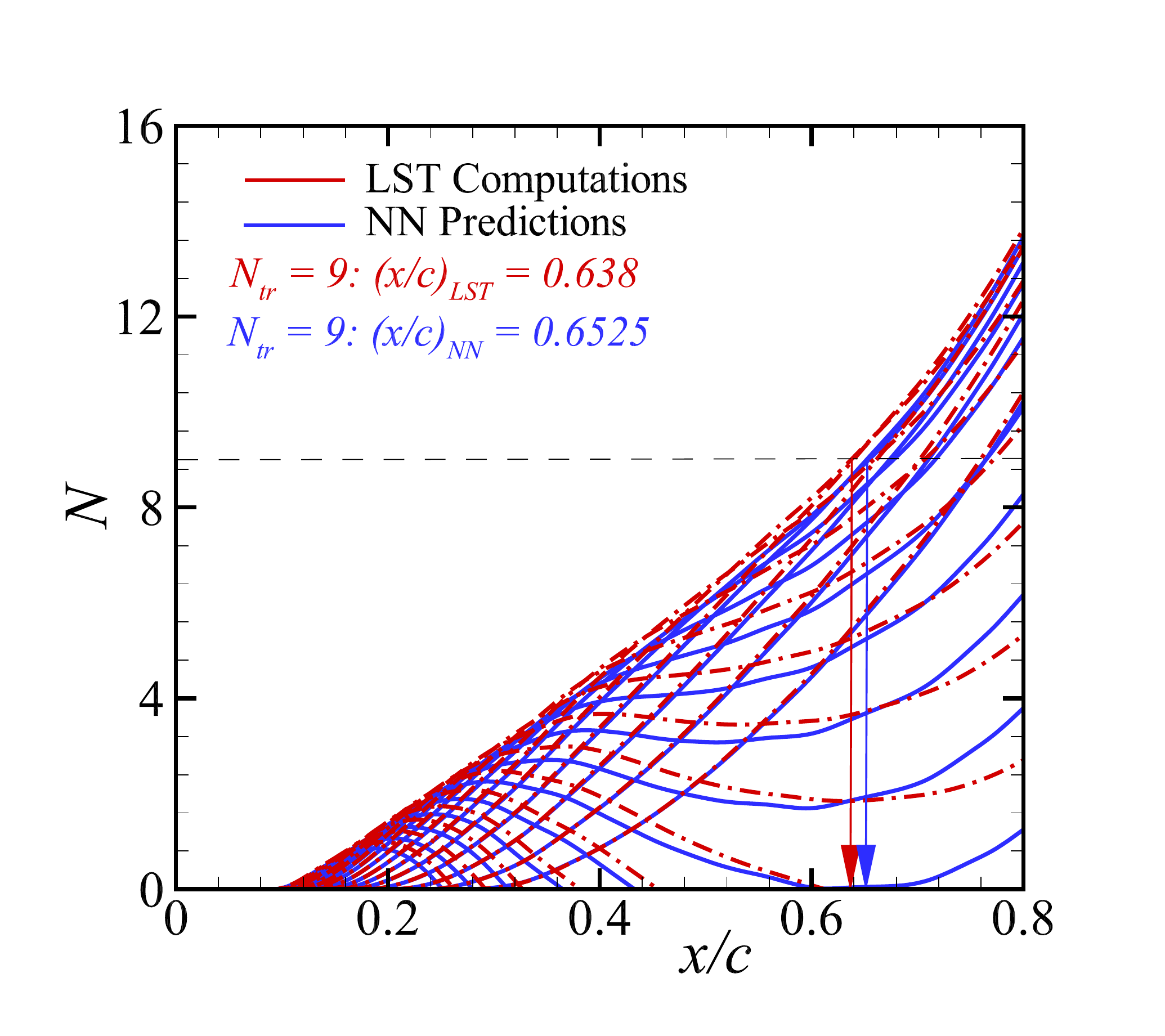}} \quad
    \subfloat[NLF-0416,  $\alpha = 0^\circ$] {\includegraphics[width=0.4\textwidth]{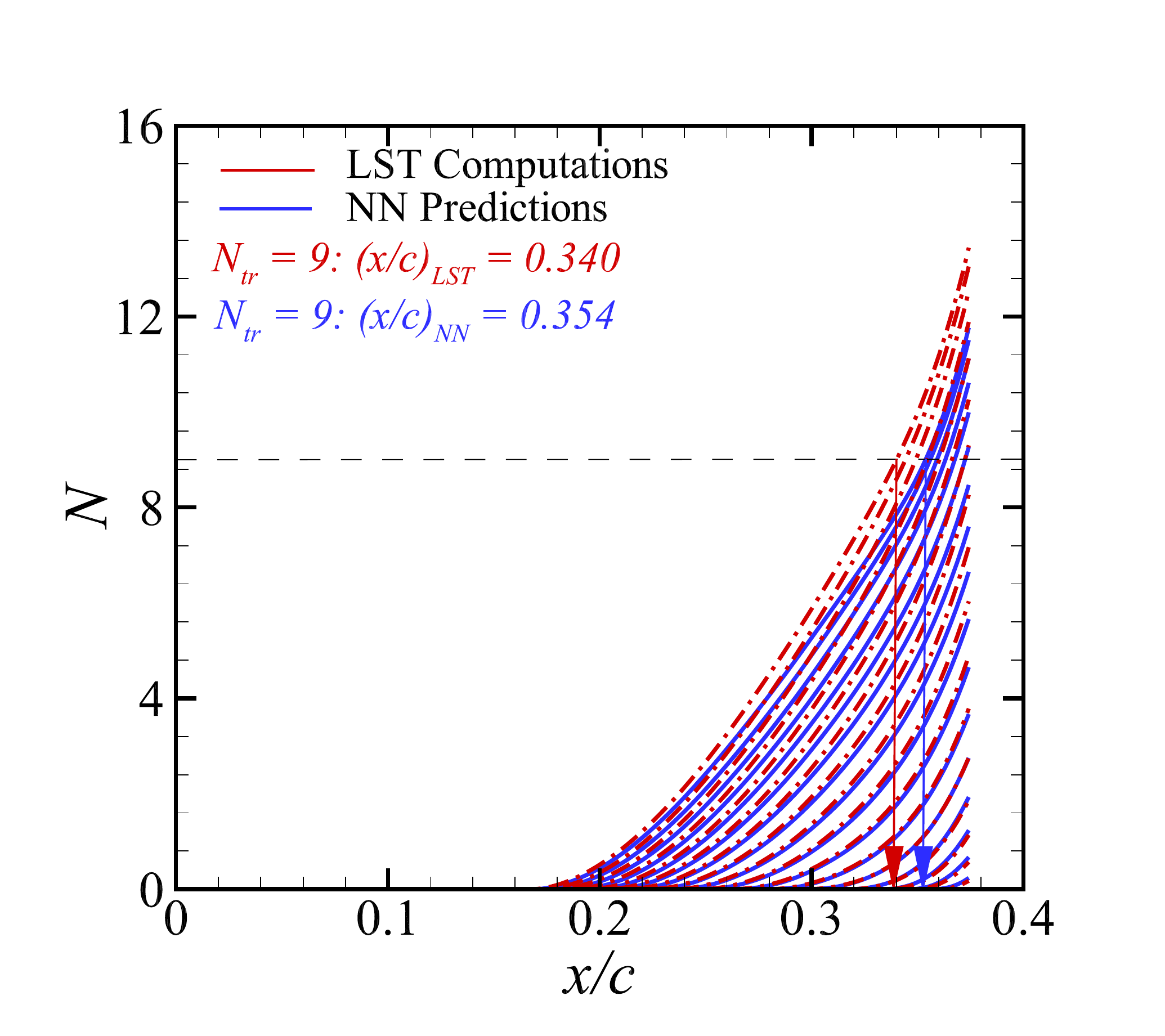}} \\
    \subfloat[NLF-0416,  $\alpha = 2^\circ$] {\includegraphics[width=0.4\textwidth]{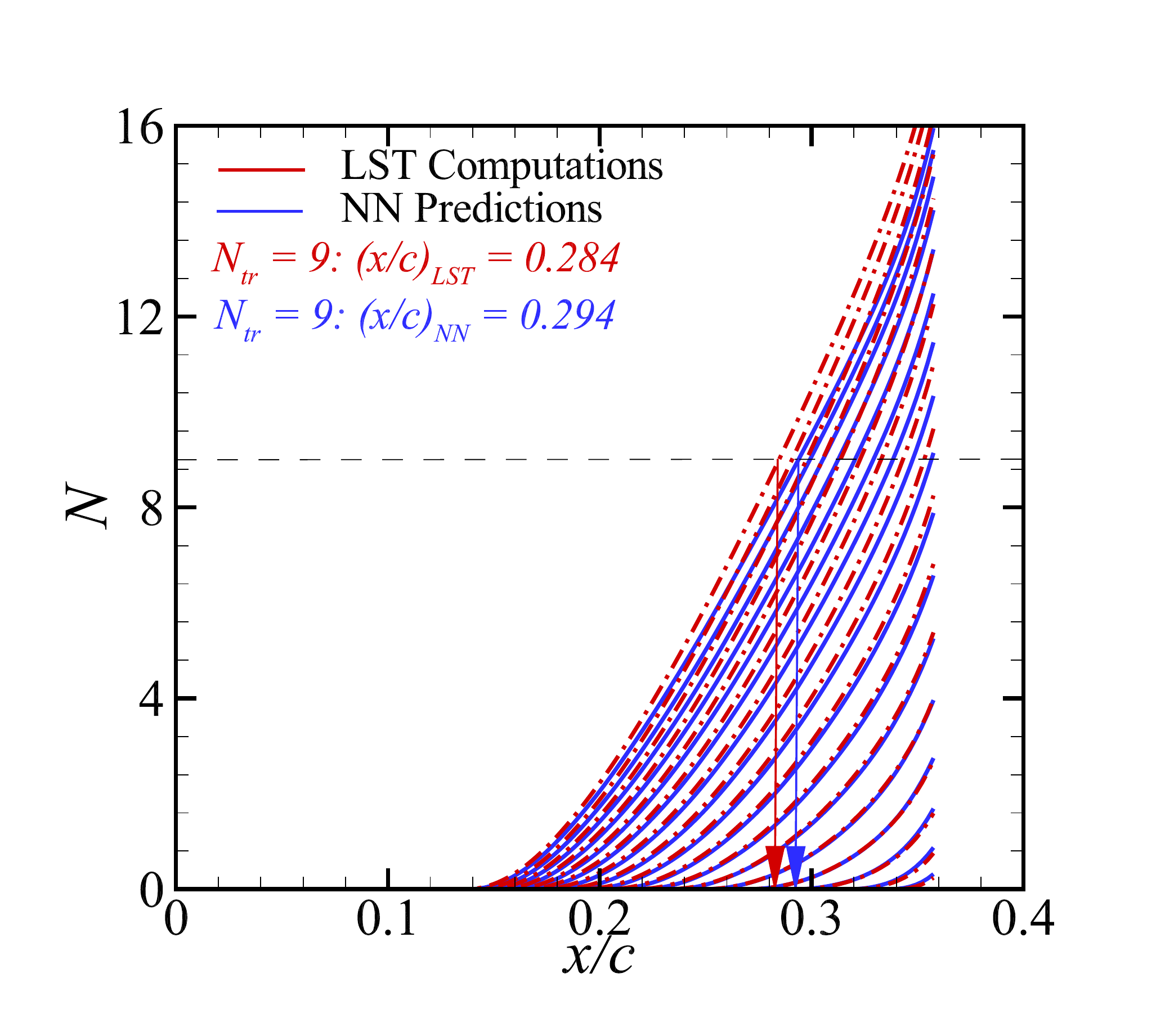}} \quad
    \subfloat[NLF-0416,  $\alpha = 5^\circ$] {\includegraphics[width=0.4\textwidth]{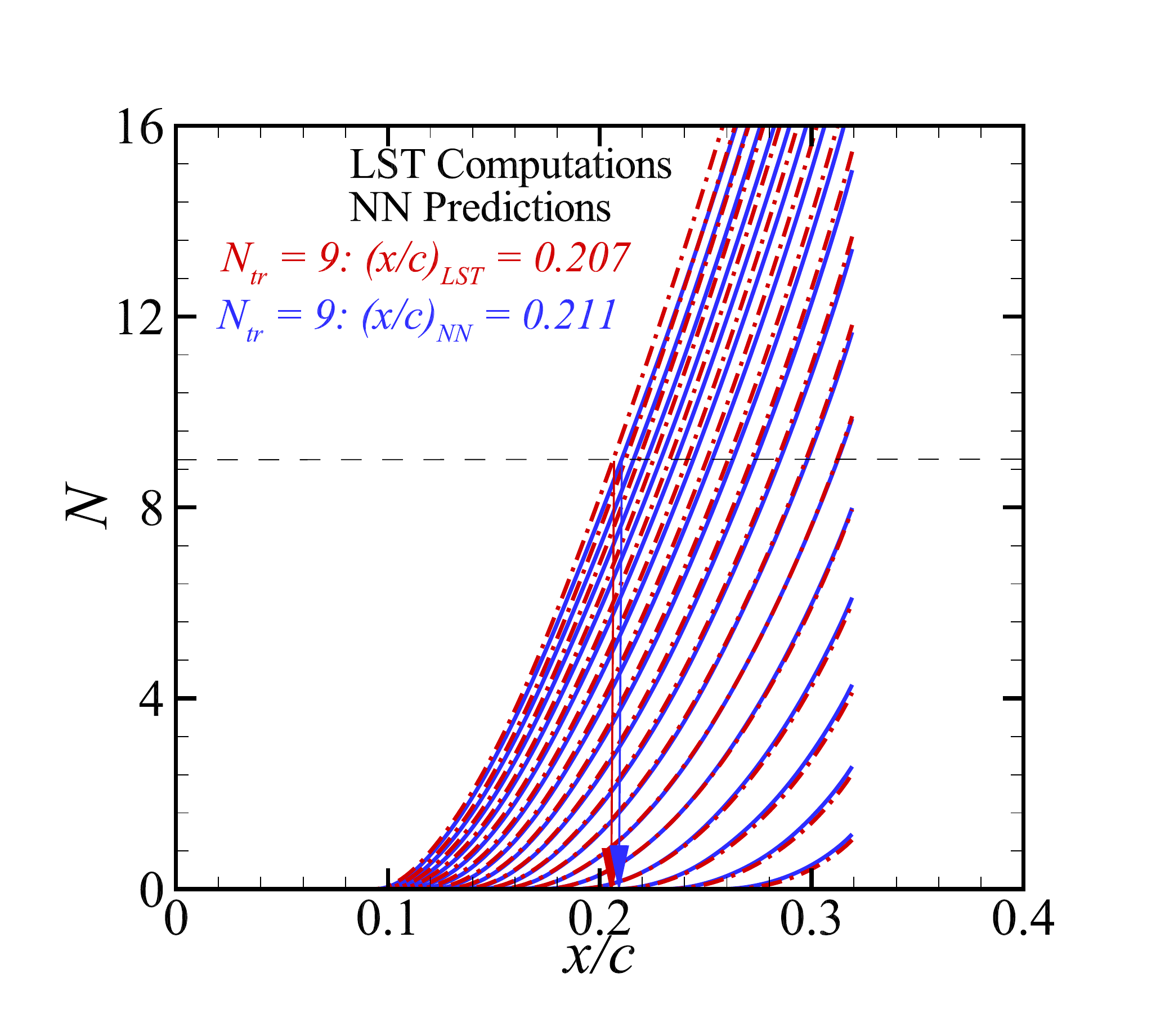}}
    \caption{N-factor curves for non-self-similar boundary layer profiles over the upper surface of a symmetric HSK airfoil section and an asymmetrical NLF-0416 airfoil at different Angles of Attack ($\alpha$). Transition location corresponds to the critical value of $N=9$ (marked by a dashed line). Corresponding transition onset locations are mentioned on the upper left corner and marked on the horizontal axis as predicted by the proposed neural network $\text{C}_1$ model (blue arrow) and computed by Linear Stability Theory (LST) (red arrow). The neural network model was trained using the Falkner Skan database (with self-similar boundary layers). Red and blue lines correspond to N-factor curves.}
    \label{fig:n_plots}
\end{figure}

As the underlying purpose of the proposed neural network model is to predict the transition onset location, we next evaluate the predictive performance of the neural network model $\text{C}_1$ for a symmetric HSK airfoil section and an asymmetric NLF-0416 airfoil section. These two examples collectively cover both favorable and adverse pressure gradients.  For this assessment, the proposed neural network model has been trained on the complete Falkner--Skan database while the testing dataset corresponds to the upper surface of the airfoil sections. Unlike the Falkner-Skan database, the boundary layer flows on these airfoils evolve in a non-self-similar manner.  Therefore, the evaluation of the neural network trained on the database of self-similar profiles allows us to gauge the practical utility of this model, i.e., the capacity to generalize the predictions to arbitrary, but still attached, boundary layer profiles. Figure~\ref{fig:n_plots} shows the corresponding results where the predicted N-factor curves for instability waves with a selected set of disturbance frequencies have been superposed on those based on the Linear Stability Theory (LST). The predicted N-factor curves are computed by integrating the local amplification rates $\sigma$ predicted by the proposed neural network model. Figure~\ref{fig:n_plots}a shows the comparison plots for the symmetric HSK airfoil section wherein the abscissa corresponds to the surface location along the upper surface of the airfoil section (scaled with respect to the airfoil chord length) and the $N$-factor values are plotted along the ordinate. As discussed in Section~\ref{sec:eN}, transition is predicted to occur when the value of $N$ reaches a critical value, which is chosen to be $N_{tr}=9$ for the purpose of this comparison~\citep{ingen2008, bushnell1989}. This critical value of N is marked by a dashed line on the plot, whereas the corresponding transition locations predicted by the neural network and the LST are indicated by blue and red arrows, respectively, and are also listed in the upper left portion of the figure. The error in the neural network prediction for the transition location based on $N_{tr}=9$ is approximately 2\% for the HSK airfoil. Similar results are shown in Figs. 3b-3d for the predicted transition location along the suction surface of the asymmetric NLF-0416 airfoil section at selected angles of attacks.  For these cases, the transition onset location has been predicted to within an error of approximately 2-5\% by the neural network model.

\subsection{Automatic, Data-Driven Feature Extraction of Boundary Layer Profiles}
The primary advantage of the proposed neural network is the feature extraction capability of the convolutional neural network (CNN) (Fig.~\ref{fig:PICNN}). The CNN is able to distill information from the boundary layer profiles in a physically consistent manner, i.e., by considering the boundary layer profiles as continuous functions and by encoding the information into a set of parameters indicated by the (green) shaded neurons $\Psi$ in Fig.~\ref{fig:PICNN}b. The CNN provides a mapping from the space of boundary layer profiles to the physical parameter space in an automated, data-driven fashion, i.e., without requiring the user to specify an explicitly defined learning target for the CNN. To demonstrate this capability of the proposed neural network, we consider the case where the CNN maps the distilled information from the boundary layer profile $U_j (j = 1, 2, …, 41)$ to a single parameter $\Psi$ at the interface between the CNN and the fully connected network in Fig. ~\ref{fig:PICNN}. In essence, this process mimics the behavior of the fully-connected network A from Table II by choosing the CNN parameters to encode a single feature from the boundary layer profiles. Because the CNN does not make any prior assumptions about what this single feature should be, one might expect that the predictive performance of the CNN architecture $\text{C}_1^*$ would be better than that of the fully connected network $A$.  Somewhat surprisingly, however, comparison of the respective testing errors for the asymmetrical NLF-0416 airfoil indicates that the network $A$ performs slightly better than $\text{C}_1^*$ (testing error of 21.2\% versus 24.5\%). The explanation of this relative performance is left as a topic for future studies.  However, it does seem to provide independent evidence that supports the practice of using the analytically defined shape factor $H$ as a nearly optimal scalar representation of the boundary layer profile for the purpose of predicting the amplification rates. Next, we evaluated the parameter $\Psi$ for each of the 111 boundary layer profiles along the upper surface of the asymmetrical NLF-0416 airfoil by using the convolutional neural network model that was trained on the full Falkner-Skan database.  For convenience of interpretation, we normalize (i.e., shift and scale) the learned parameter $\Psi$ to $\Tilde{\Psi}$ such that the latter falls within the range of [0, 1].  Figure 4 indicates the variation in $\Tilde{\Psi}$  with a similarly normalized shape factor $\Tilde{H}$ of the velocity profiles that is defined in the inset of the figure.  The plot shows a nearly linear relationship between the CNN-extracted feature $\Tilde{\Psi}$ and the physically defined counterpart $\Tilde{H}$.  We point out that, because the activation functions for the layer $\Psi$ is linear, the mapping between the two parameters remains unchanged if we scale all the weights leading into these neurons by an arbitrary factor $a$ and the value of these neurons by $1/a$. As such, the normalization process is well justified.
\begin{figure}
    \centering
    \includegraphics[width=0.54\textwidth]{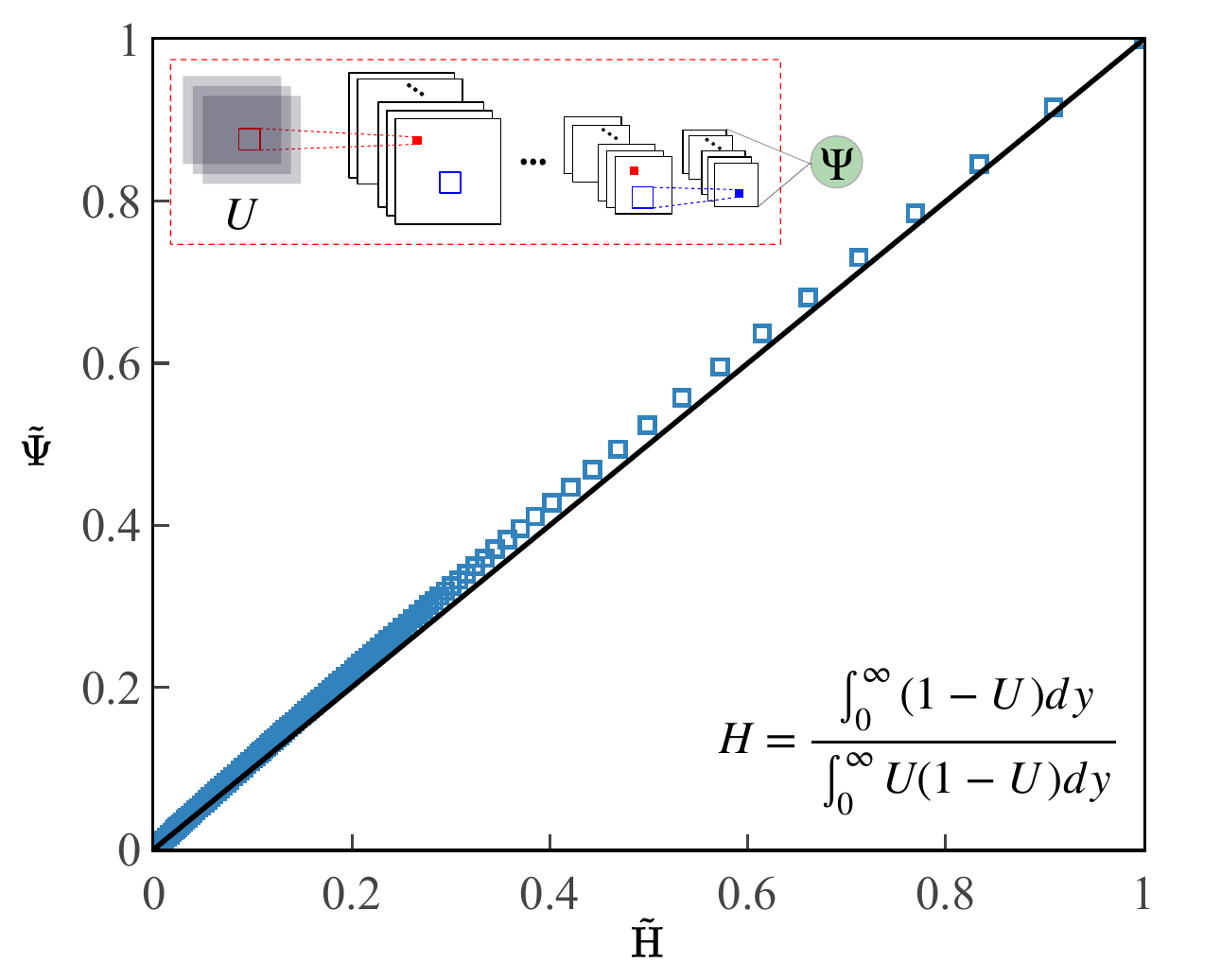}
    \caption{Correlation between normalized CNN learned parameter ($\Tilde{\Psi}$) from non-self-similar boundary layer profile $U$ and normalized shape factor ($\Tilde{H}$), at 111 locations along the upper surface of an asymmetric NLF-0416 airfoil section at $0$ degrees angle of attack. The learned parameter $\Psi$ and the shape parameter $H$ are normalized to within the range $[0, 1]$ to facilitate comparison. The Falkner--Skan database (with self-similar boundary layers) has been used for training.  \label{fig:correlation_sch}}
\end{figure}

\begin{figure}
    \centering
    \includegraphics[width=0.57\textwidth]{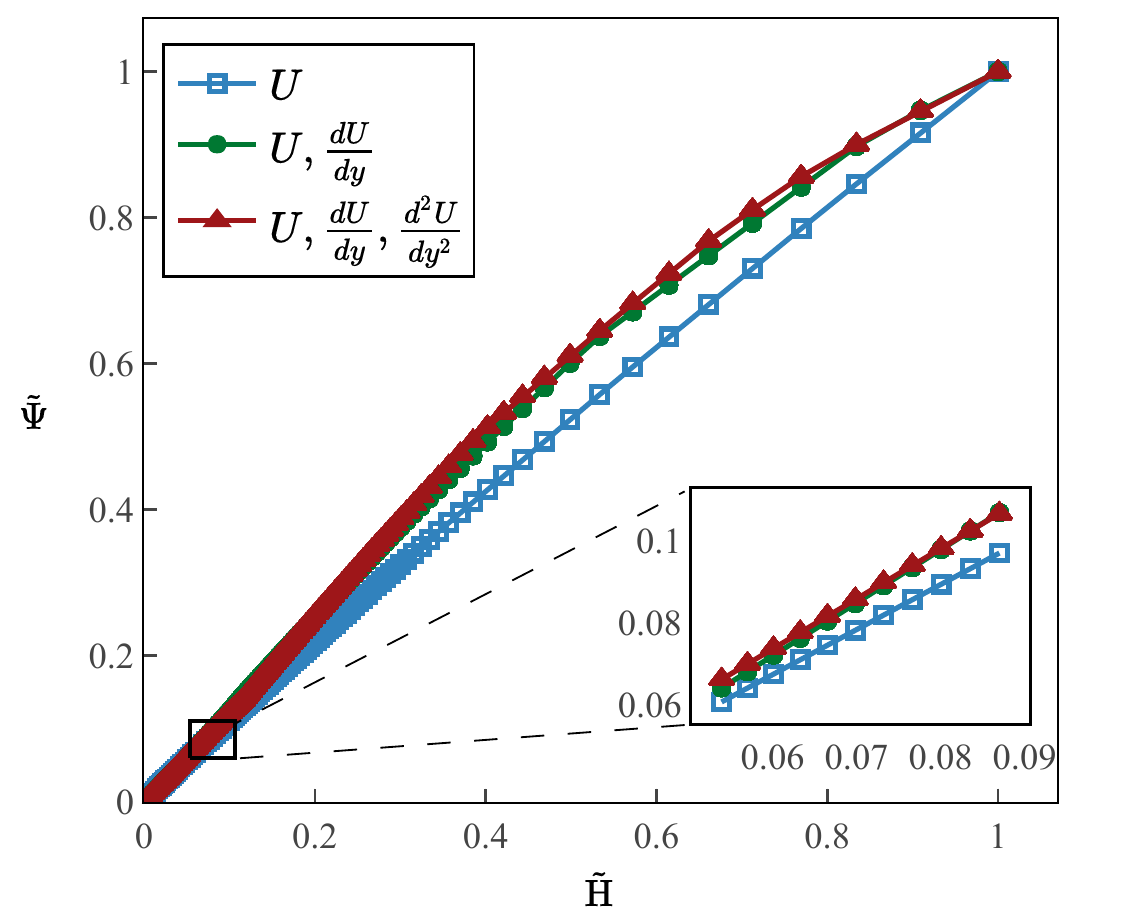}
    \caption{Comparison of correlation between normalized CNN learned parameter ($\Tilde{\Psi}$) for varying sets of non-self-similar boundary layer profiles (corresponding to networks $\text{C}_1^*$,$\text{C}_2^*$ and $\text{C}_3^*$) and normalized shape factor ($\Tilde{H}$), at 111 locations along the upper surface of an asymmetric NLF-0416 airfoil section at $0$ degrees angle of attack. Falkner--Skan database (with self-similar boundary layers) has been used for training. \label{fig:correlation}}
\end{figure}

A similar evaluation of the data-driven feature extraction capability of the CNN was performed by varying the number of boundary layer profiles used as input to the neural network.  Specifically, the correlation analysis from Fig. 4 was repeated by including the first and second derivatives of the velocity profile in addition to the velocity profile itself. The results of this analysis for all three networks (namely, network $\text{C}_1^*$ with $U$, $dU/dy$, and $d^2U/dy^2$ profiles as input, $\text{C}_2^*$ with $U$ and $dU/dy$ as input, and $\text{C}_3^*$ with $U$ only). The results are presented and compared in Fig.~\ref{fig:correlation}. It can be observed that the correlation between $\Tilde{\Psi}$ and $\Tilde{H}$ is linear for $\text{C}_3^*$, where only the velocity profile $U$ is used as input. In comparison, a mild nonlinearity may be observed in the $\Tilde{\Psi}$-$\Tilde{H}$  relations for the $\text{C}_2^*$ and $\text{C}_1^*$ networks, where the first and second derivatives $dU/dy$ and $d^2U/dy^2$ of the velocity are also introduced as additional input features to the CNN.  However, whether the relation is linear or nonlinear does not have any major consequence for our purpose.  What is more important is that all three networks exhibit a one-to-one correspondence between the CNN-extracted quantity $\Psi$ from the boundary layer profiles and the physically defined quantity $\Tilde{H}$. Thus, we may conclude that the proposed CNN architecture has a robust performance in goal-oriented feature extraction.  In particular, without any explicit instruction from the user, the CNN has been able to encode the boundary layer profiles (i.e., velocity and its derivatives) into a quantity $\Psi$ that is predictive of the amplification rate $\sigma$ when used in conjunction with the other physical parameters $\omega$ and $Re_\theta$. The quantity $\Psi$ is strongly correlated to the shape factor $H$, which is known to be correlated with the stability characteristics~\citep{drela2018}. This observation points to the physically consistent nature of the proposed neural network architecture. In addition to the theoretical importance, the results also have practical importance. Specifically, the results above highlight the intrinsic potential of the proposed network to allow additional features of the boundary layer profiles to be taken into account with manageable computational costs and without compromising the robustness of the network performance.

Admittedly, the shape factor H of the Falkner-Skan boundary layers can be easily evaluated and then used to map the complex dependence of the disturbance amplification rate on the underlying mean flow through a fully connected neural network (Fig.~\ref{fig:FCNN}a). However, in several other cases, such as high-speed flows over blunt nose configurations~\citep{paredes2020} or flows where the edge of the boundary layer cannot be easily determined, the shape factor H cannot be defined in a consistent and accurate manner and/or computed straightforwardly. In such flows, the proposed convolutional neural network provides a more general and effective architecture for modelling the local instability amplification rates based on the boundary layer profiles and other relevant physical parameters. Although a reduced representation of the input features can also be achieved via dimensionality reduction techniques, such as principal  component analysis~\citep{rajnarayan2013}, the proposed neural network architecture provides an easier technique to encode the targeted information from boundary layer profiles into a smaller set of parameters. This feature extraction capability of the CNN is likely to assume an even greater significance for three dimensional boundary layer profiles involving the additional crossflow velocity component, and high-speed flows that involve the profiles of thermodynamic quantities such as the density and/or temperature, and finally, for boundary layer flows that are inhomogeneous in two spatial coordinates instead of just the wall-normal coordinate, e.g., planar boundary layer  profiles that vary along both the wall-normal and the spanwise directions.

\subsection{Comparison to a Fully Connected Network with profile inputs}
\begin{figure}
    \centering
    \subfloat[Validation plot, growth rate predictions for CNN $\text{C}_1$]{\includegraphics[width=0.38\textwidth]{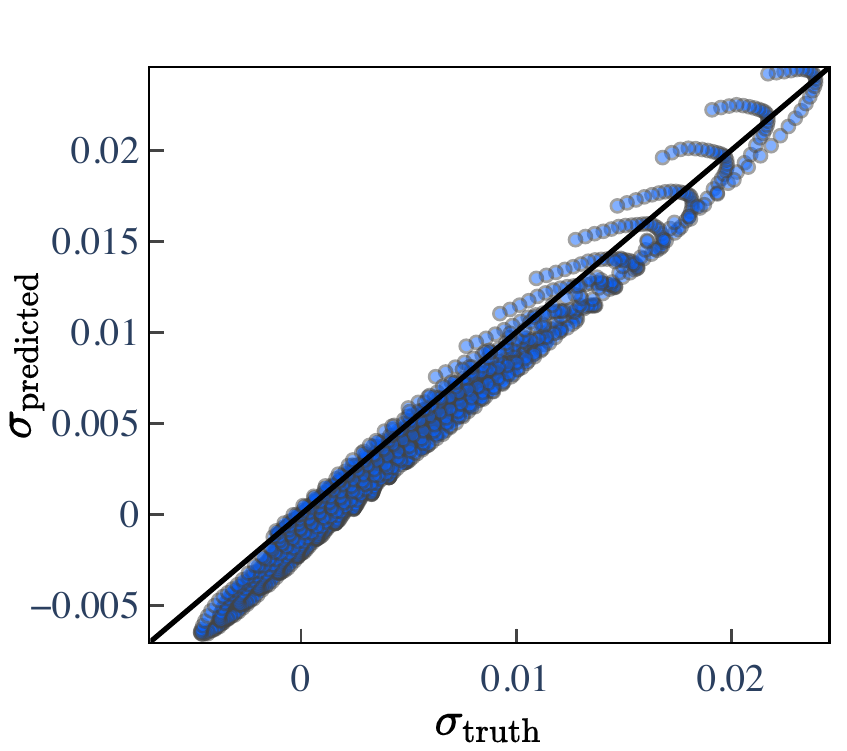}} \quad
    \subfloat[N-factor curves, CNN $\text{C}_1$]{\includegraphics[width=0.39\textwidth]{LST_nlf0416_aoa0_upper_Re9M_prof_cnn_Nfactors-eps-converted-to.pdf}} \\
    \subfloat[Validation plot, growth rate predictions for fully connected network $\text{B}$]{\includegraphics[width=0.38\textwidth]{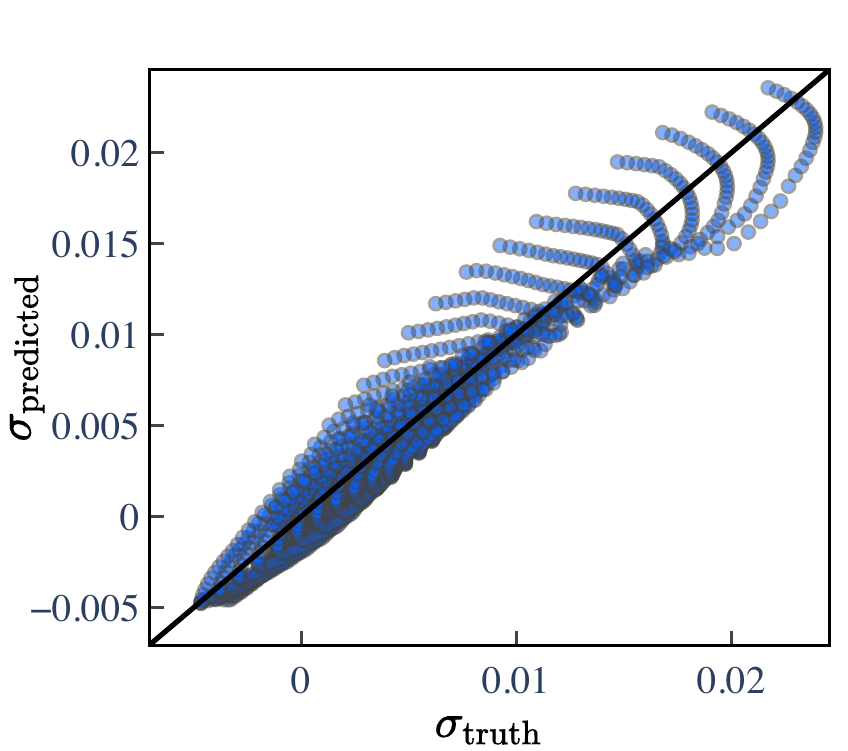}} \quad
    \subfloat[N-factor curves, fully connected network $\text{B}$  ]{\includegraphics[width=0.39\textwidth]{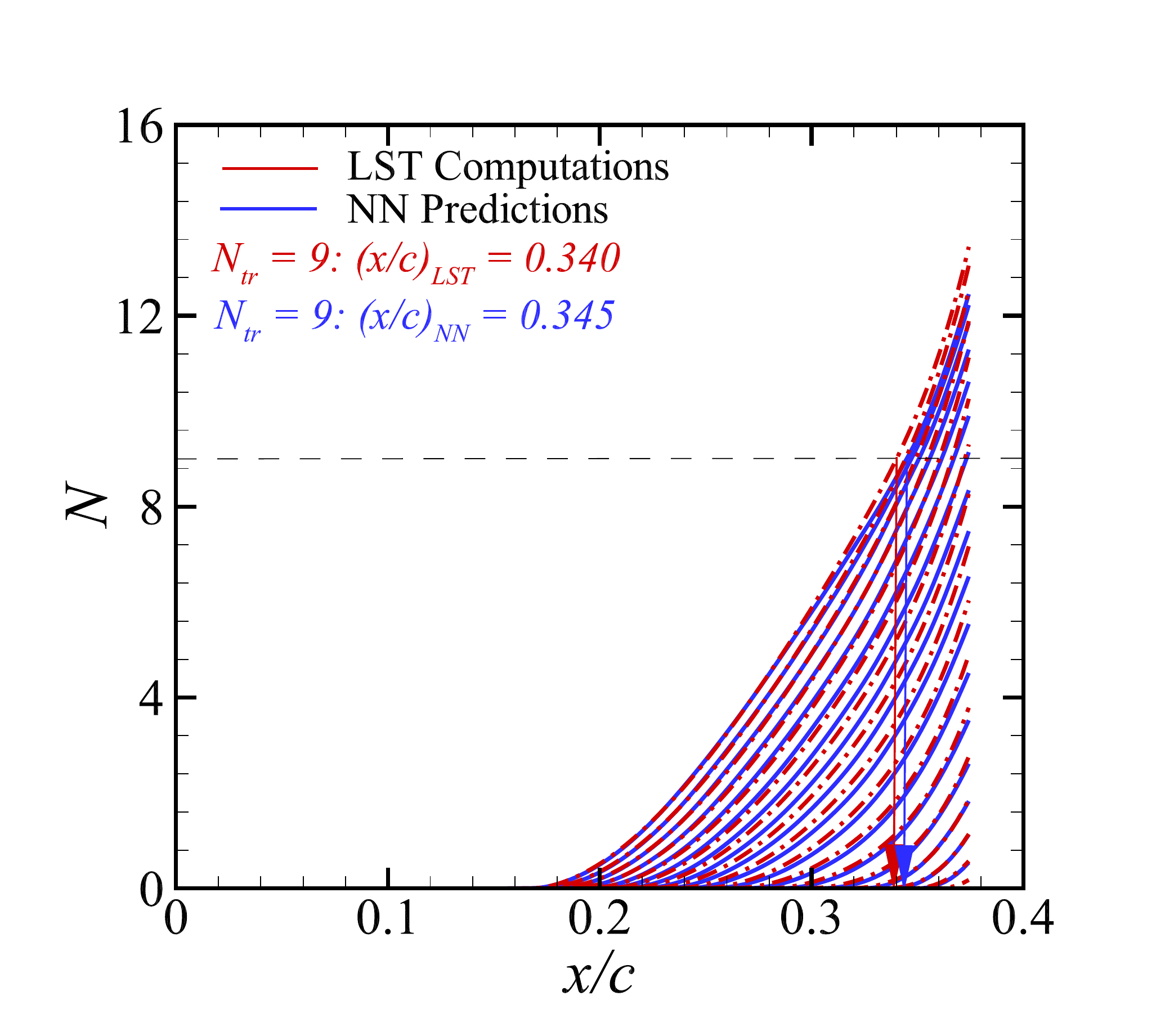}}
    \caption{Comparison between fully connected network $\text{B}$ and proposed convolutional neural network $\text{C}_1$. Validation plots of local instability amplification rates $\mathbf{\sigma}$ and N-factor curves for an asymmetric NLF-0416 airfoil section at $0$ degrees angle of attack are given for both networks. Transition location corresponds to the critical value of $N=9$ (marked by dashed line). Corresponding transition onset locations are mentioned below the legends in (b) and (d) and marked on the horizontal axis as predicted by neural network model (NN) (blue arrow) and computed by Linear Stability Theory (LST) (red arrow).}
    \label{fig:hsk_plots}
\end{figure}

The proposed convolutional neural network has also been assessed against the straightforward method of directly introducing the full boundary layer profiles as input features to a fully connected neural network, as shown in Fig.~\ref{fig:FCNN}b.   The latter architecture may be viewed as a generalization of the architecture in Ref.~\citep{crouch2002}.  Figure~\ref{fig:hsk_plots} shows a comparison between the predictive performances of the proposed convolutional neural network (Fig.~\ref{fig:PICNN}) and the fully connected network (Fig.~\ref{fig:FCNN}b). Both neural network models have been trained using the stability database for Falkner-Skan profiles and the predictive performance is evaluated for instability amplification over the upper surface of the HSK airfoil section. To help ensure a fair comparison, the number of model parameters for both models is kept approximately equal as given in Table I for networks $\text{C}_1$ and $\text{B}$, respectively. The predicted transition onset location based on each model is marked by an arrow on the x-axis and mentioned below the legends in Figs.~\ref{fig:hsk_plots}b and~\ref{fig:hsk_plots}d.  For reference, the predictions based on direct stability computations are also shown.  Although the validation plots for instability amplification rates in Figs.~\ref{fig:hsk_plots}a and~\ref{fig:hsk_plots}c show better predictive performance for the convolutional neural network, the prediction of transition onset location is better for the fully connected network. The transition location predicted by the fully connected network is within 1.1\% of the transition onset location based on the linear stability theory, whereas the convolutional neural network predicts the same with a 2.3\% error. Even though both measures of error are rather small, the qualitative trend is somewhat unexpected and requires further investigation in future studies. 

\begin{figure}
    \centering
    \includegraphics[width=0.45\textwidth]{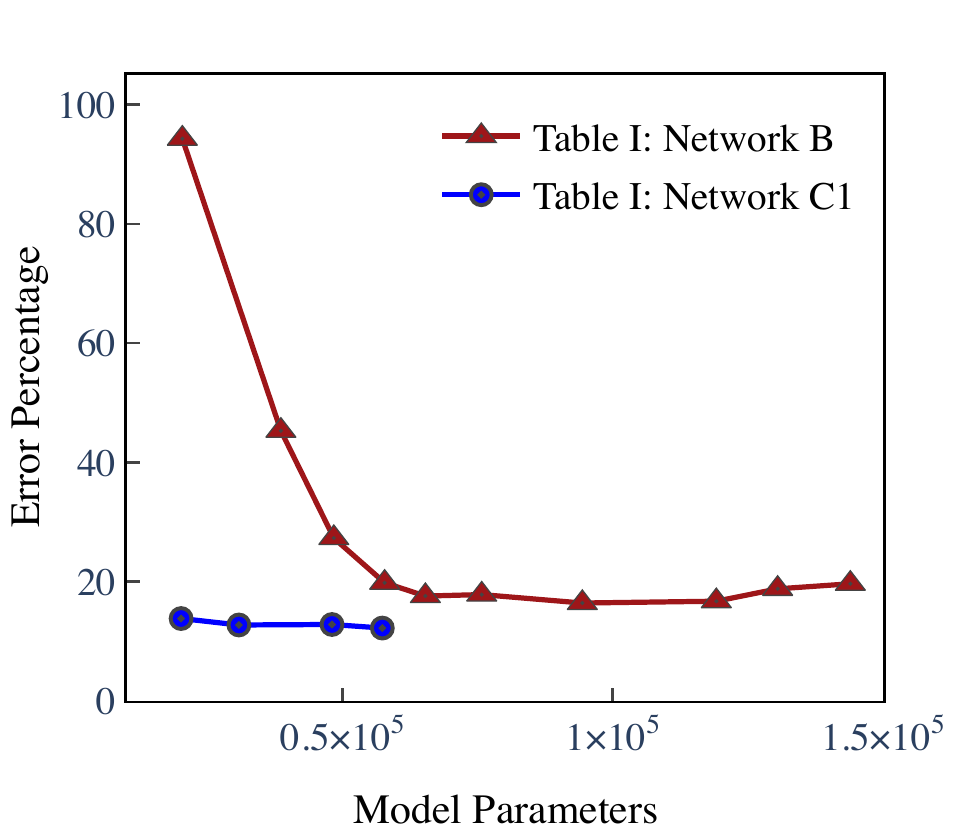}
    \caption{Comparison of testing error for the fully connected neural network (FC NN) (Fig.~\ref{fig:FCNN}b, Network B from Table I) and the convolutional neural network (Fig.~\ref{fig:PICNN}, Network $\text{C}_1$ from Table I) as a function of the total number of learnable model parameters. Input features for both models include boundary layer profiles. The Falkner--Skan database (with self-similar boundary layers) was used for training while the testing error was evaluated for an asymmetric NLF-0416 airfoil section with non-self-similar boundary layers at $0$ degrees angle of attack.}\label{fig:error_vs_para}
\end{figure}

One significant advantage of the proposed convolutional neural network (Fig.~\ref{fig:PICNN}) over the fully connected network (Fig.~\ref{fig:FCNN}b) pertains to the number of learnable model parameters required to achieve a comparable performance. With full information available as input features, a fully connected network would require a significantly higher number of model parameters, and consequently, a higher training cost to provide a comparable performance, whereas the convolutional neural network is likely to provide more robust predictive performance with a smaller number of model parameters.  Figure~\ref{fig:error_vs_para} presents the results of the analysis designed to verify this behavior.  Here, the test error percentage has been plotted against the size of the neural network model as measured by the total number of trainable model parameters. We observe that the performance of the fully connected neural network deteriorates significantly as the number of model parameters is reduced while the proposed convolution-based neural network is able to maintain its predictive performance. The convolutional neural network in the proposed model encodes the information from the boundary layer profiles into a significantly smaller number of scalar parameters (8 parameters for network $\text{C}_3$) before folding them into the ensuing fully connected portion of the overall network, along with the other physical parameters (namely, frequency of the instability wave $\omega$ and the local Reynolds number $Re_\theta$). Such feature engineering for the boundary layer profiles via the convolutional neural network enables one to achieve a comparable performance with a significantly reduced number of model parameters and training cost. 

\section{Conclusion}
\label{sec:conclude}
A neural-network-based transition model has been presented that is capable of accurately predicting the transition onset location for incompressible, two-dimensional attached flows in a physically informed manner without requiring the direct computations using the linear stability theory. The proposed model has the ability to encode information, using convolutional layers, from boundary layer profiles (velocity and its derivatives)  into a set of integral quantities. More importantly, the encoded feature $\tilde{\Psi}$ shows strong correlation with, or even one-to-one correspondence to the physically defined shape parameter $H$, which clearly demonstrates the physically consistent nature of the proposed neural network. These encoded integral quantities are then nonlinearly mapped to local instability amplification rate $\sigma$ along with other scalar disturbance characteristics ($\omega$ and $Re_{\theta}$). The proposed model is shown to have robust predictive performance, clear physical interpretation, and superior computational efficiency.

The CNN architecture presented herein can be easily generalized to other instability mechanisms and, in follow-on (and as yet unpublished work), we have demonstrated the application of this architecture to second mode instabilities in high-speed boundary layers that cannot be predicted well on the basis of local shape factors of the boundary layer profiles. Thus, it could become the means for physics-based transition prediction in practical applications of computational fluid dynamics codes.

\section*{Acknowledgments}
The computational resources used for this project were provided by the Advanced Research Computing (ARC) of Virginia Tech, which is gratefully acknowledged.

\end{document}